\documentclass{elsart}

\usepackage{amssymb}
\usepackage{amsfonts}
\usepackage{amsmath}

\begin{document}

\begin{frontmatter}

\rightline{STUPP-08-196}

\title{Standard(-like) Model from an SO(12) Grand Unified Theory in six-dimensions with $S_2$ extra-space }

\author[Saitama]{Takaaki Nomura \corauthref{cor}} and
\corauth[cor]{Corresponding author. 
telephon number: 810488589102}
\ead{nomura@krishna.th.phy.saitama-u.ac.jp}
\author[Saitama]{Joe Sato} 
\ead{joe@phy.saitama-u.ac.jp}

\address[Saitama]{Department of Physics,
  Saitama University,
  Shimo-Okubo,
  Sakura-ku,
  Saitama 355-8570, Japan}

\begin{abstract}
We analyze a gauge-Higgs unification model which is based on 
a gauge theory defined on a six-dimensional spacetime with an $S^2$ extra-space.
We impose a symmetry condition for a gauge field and non-trivial boundary conditions of the $S^2$.
We provide the scheme for constructing a four-dimensional theory from the six-dimensional 
gauge theory under these conditions.
We then construct a concrete model based on an SO(12) gauge theory with fermions which lie in a 32 
representation of SO(12), under the scheme.
This model leads to a Standard-Model(-like) gauge theory which has gauge symmetry 
SU(3) $\times$ SU(2)$_L$ $\times$ U(1)$_Y$($\times$ U(1)$^2$) and 
one generation of SM fermions, in four-dimensions.
The Higgs sector of the model is also analyzed, and it is shown that the electroweak symmetry breaking 
and the prediction of W-boson and Higgs-boson masses are obtained.
\end{abstract}


\end{frontmatter}
\section{Introduction}
The Higgs sector of the Standard Model (SM) plays an essential role in the mechanism of 
spontaneous breaking of the gauge symmetry from SU(3)$_C$ $\times$ SU(2)$_L$ $\times$ U(1)$_Y$ 
down to SU(3)$_C$ $\times$ U(1)$_{EM}$, giving masses to the elementary particles. 
The SM, however, does not address even the most fundamental nature of the Higgs sector, 
such as 
the mass of Higgs particles and the Higgs self-coupling constant.
Thus the Higgs sector is not only the last frontier of the SM, but it will also provide the 
key clue to the physics beyond the SM.


The gauge-Higgs unification is one of 
the attractive approaches to the physics beyond the SM in this regard~\cite{%
  Manton:1979kb,%
  Fairlie:1979at,%
  Fairlie:1979zy%
} (for recent approaches, 
 see Refs.~\cite{%
  Hall:2001zb,%
  Burdman:2002se,%
  Gogoladze:2003ci,%
  Scrucca:2003ut,%
  Haba:2004qf,%
  Biggio:2004kr,%
  Maru:2004,%
  Haba:2005kc,%
  Hosotani:2006qp,%
  Sakamoto:2006wf,%
  Maru:2006,%
  Hosotani:2007qw,%
  Sakamura:2007qz,%
  Medina:2007hz,%
  maru2007,%
  Adachi:2007tj,%
  Gogoladze:2007ey%
}). \\
In this approach, the Higgs particles originate from the extra-dimensional components of the gauge field 
of a gauge theory defined on spacetime with dimensions larger than four.
Thus the Higgs sector is embraced into the gauge interactions in the higher-dimensional spacetime 
and part of the fundamental properties of Higgs scalar is determined from the gauge interactions.
%


We consider a gauge-Higgs unification model based on a gauge theory 
as defined on the six-dimensional spacetime with 
the extra-space which has the structure of two-sphere $S^2$.
We can impose on the fields of this gauge theory the symmetry condition which identifies the gauge transformation 
as the isometry transformation of $S^2$ as in the coset space dimensional reduction(CSDR) 
scheme~\cite{Manton:1979kb,Forgacs:1979zs,Kapetanakis:1992hf,Chatzistavrakidis:2007by,Zoupanos08}
, since the $S^2$ has the coset space structure such as $S^2$=SU(2)/U(1).
We then impose on the gauge field the symmetry in order to carry out the dimensional reduction of the 
gauge sector.
The dimensional reduction is explicitly carried out by applying the solution of the symmetry condition, and
a background gauge field is introduced as a part of 
the solution of the symmetry condition~\cite{Manton:1979kb}.
We obtain, by the dimensional reduction, the scalar sector with a potential term which leads 
to spontaneous symmetry breaking. 
The symmetry also restricts the gauge symmetry and the scalar contents originated from extra guage 
field components
 in four-dimensions.
We, however, do not impose the symmetry on 
the fermion of the gauge theory, in contrast to other CSDR models. 
We then have massive Kaluza-Klein(KK) modes of fermion in four-dimensions while gauge and scalar fields have no 
massive KK mode, and would obtain a dark-matter candidate.
Generally, the KK modes do not have massless mode because of positive curvature of $S^2$~\cite{A.A.Abrikosov}.
We, however, obtain a massless KK mode because of existence of background gauge field;
the fermion components which have the massless mode are determined by the background gauge field.


Gauge theories with the symmetry condition are well investigated to construct a model which provide
Grand Unified Theory (GUT) 
in four-dimensions~\cite{Kapetanakis:1992hf,Chapline:1982wy,10dim-Model:K12,10dim-Model:D14,10dim-Model:B,CSDR14D}.
No known model, however, reproduced the full particle contents of GUTs.
We generally cannot obtain the Higgs particles which properly break a GUT gauge symmetry, while 
one or more generation of fermions and SM Higgs-doublet could be obtained. 
We then impose on fields of a six-dimensional theory the non-trivial boundary conditions of $S^2$ 
together with the symmetry condition in order to overcome the difficulty.
A GUT gauge symmetry can be broken to SM gauge symmetry by the non-trivial boundary conditions
(for cases with orbifold extra-space, 
see for example~\cite{%
  Hall:2001zb,%
  Burdman:2002se,%
  Gogoladze:2003ci,%
  Scrucca:2003ut,%
  Haba:2004qf,%
  Haba:2005kc,%
  Hosotani:2006qp,%
  Sakamura:2007qz,%
  Medina:2007hz,%
  maru2007,%
  Kawamura1,%
  Kawamura2}).


In this paper, we analyze 
the gauge theory defined on the six-dimensional spacetime which has 
$S^2$ as extra-space, with the symmetry condition and non-trivial boundary conditions.
The gauge symmetry, scalar contents and massless fermion contents are determined by the symmetry condition and the 
boundary conditions.
First, we provide the scheme for constructing a four-dimensional theory from the six-dimensional gauge theory.
We then construct the model based on SO(12) gauge symmetry and show that SM-Higgs doublet and one generation of 
massless fermions are obtained in four-dimensions.
We also find that the electroweak symmetry breaking is realized 
and Higgs mass value is predicted
 by analyzing Higgs sector of the model.


This paper is organized as follows.
In sec.~\ref{CSDR}, we give the scheme for constructing a four-dimensional theory from a gauge theory on 
six-dimensional spacetime which has extra space as two-sphere $S^2$ with the symmetry condition and 
non-trivial boundary conditions.
In sec.~\ref{SO(12)model}, we construct the model based on SO(12) gauge symmetry.
We summarize our results in sec.~\ref{summary}.



\section{Six-dimensional gauge theory with extra-space $S^2$ under the symmetry condition and non-trivial boundary conditions }
\label{CSDR}

In this section, we develop the scheme for constructing a four-dimensional theory from a gauge theory on 
six-dimensional spacetime which has extra-space as two-sphere $S^2$ with the symmetry condition and 
non-trivial boundary conditions.


\subsection{A Gauge theory on six-dimensional spacetime with $S_2$ extra-space}

We begin with a gauge theory with a gauge group $G$ defined on a six-dimensional spacetime $M^6$.
The spacetime $M^6$ is assumed to be a direct product of the four-dimensional 
Minkowski spacetime $M^4$ and two-sphere $S^2$ such that $M^6=M^4 \times S^2$.
The two-sphere $S^2$ is a unique two-dimensional coset space, and can be written 
as $S^2 = \mathrm{SU}(2)_I/\mathrm{U}(1)_I$, where U(1)$_I$ is the subgroup of SU(2)$_I$.
This coset space structure of $S^2$ requires 
that $S^2$ has the isometry group SU(2)$_I$, and that the group U(1)$_I$ is embedded into 
the group SO(2) which is a subgroup of the Lorentz group SO(1,5).
We denote the coordinate of $M^6$ by $X^M=(x^{\mu},y^{\theta}=\theta,y^{\phi}=\phi)$,
where $x^{\mu}$ and $\{ \theta,\phi \}$ are $M^4$ coordinates and $S^2$ spherical coordinates,
 respectively.
The spacetime index $M$ runs over $\mu$ $\in$ $\{ 0,1,2,3 \}$ and $\alpha$ $\in$ $\{ \theta,\phi \}$.
The metric of $M^6$, denoted by $g_{MN}$, can be written as
\begin{equation}
g_{MN} = \begin{pmatrix} \eta_{\mu \nu} & 0 \\ 0 & -g_{\alpha \beta} \end{pmatrix}, 
\end{equation}
where $\eta_{\mu \nu}= diag(1,-1,-1,-1)$ and $g_{\alpha \beta}= diag(1, \sin^{-2} \theta)$ 
are metric of $M^4$ and $S^2$ respectively.
Notice that we omit the radius $R$ of $S^2$ in this discussion. 
We define the vielbein $e^{M}_{A}$ 
that connects the metric of $M^6$ and that of 
the tangent space of $M^6$, denoted by $h_{AB}$, as $g_{MN}=e_M^{A} e_N^B h_{AB}$. 
Here $A=(\mu,a)$, where $a$ $\in$ $\{ 4,5 \}$, is the index for the coordinates of tangent space of $M^6$. 
The explicit form of the vielbeins are summarized in the Appendix.
We introduce a gauge field $A_{M}(x,y)=(A_{\mu}(x,y),A_{\alpha}(x,y))$, which belongs to 
the adjoint representation of the gauge group $G$, and fermions $\psi(x,y)$, which lies in 
a representation $F$ of $G$.
The action of this theory is given by 
\begin{equation}
\label{6Daction}
S = \int dx^4 \sin \theta d \theta d \phi \bigl(
\bar{\psi} i \Gamma^{\mu} D_{\mu} \psi + \bar{\psi} i \Gamma^{a} e^{\alpha}_{a} D_{\alpha} \psi 
- \frac{1}{4 g^2} g^{MN} g^{KL} Tr[F_{MK} F_{NL}] \bigr) ,
\end{equation}
where $F_{MN}= \partial_M A_N(X) -\partial_N A_M(X) -[A_M(X),A_N(X)]$ is the field strength, 
$D_M$ is the covariant derivative including spin connection, and $\Gamma_A$ represents 
the 6-dimensional 
Clifford algebra.
Here $D_M$ and $\Gamma_A$ can be written explicitly as,
\begin{align}
D_{\mu} &= \partial_{\mu} - A_{\mu}, \\
D_{\theta} &= \partial_{\theta} - A_{\theta}, \\
D_{\phi} &= \partial_{\phi} -i \frac{\Sigma_3}{2} \cos \theta -A_{\phi}, \\
\Gamma_{\mu} &= \gamma_{\mu} \otimes \mathbf{I}_2, \\
\Gamma_4 &= \gamma_{5} \otimes \sigma_1, \\
\Gamma_5 &= \gamma_{5} \otimes \sigma_2, 
\end{align}
where $ \{ \gamma_{\mu}, \gamma_{5} \} $ are the 4-dimensional Dirac matrices, 
$\sigma_i(i=1,2,3)$ are Pauli matrices, $\mathbf{I}_d$ is $d \times d$ identity, 
and 
$\Sigma_3$ is defined as $\Sigma_3=\mathbf{I}_4 \otimes \sigma_3$.
%



\subsection{The symmetry condition and the boundary conditions}
 
We impose on the gauge field $A_M(X)$   
the symmetry which connects SU(2)$_I$ isometry transformation on $S^2$
and the gauge transformation on the fields in order to carry out dimensional reduction, and 
the non-trivial boundary conditions of $S^2$ to restrict four-dimensional theory.
The symmetry requires that the SU(2)$_I$ coordinate transformation should be compensated by
a gauge transformation~\cite{Manton:1979kb,Forgacs:1979zs}.
The symmetry further leads to the following set of the symmetry condition on the fields:
\begin{align}
\label{symm-con-vec4} 
\xi_i^{\beta} \partial_{\beta} A_{\mu} 
&= \partial_{\alpha} W_i + [W_i,A_{\mu}],   \\
\label{symm-con-vecex}
\xi_i^{\beta} \partial_{\beta} A_{\alpha} + \partial_{\alpha} \xi_i^{\beta} A_{\beta} 
&= \partial_{\alpha} W_i + [W_i,A_{\alpha}],    
\end{align}
where $\xi_i^{\alpha}$ is the Killing vectors generating SU(2)$_I$ symmetry and $W_i$ are some fields which 
generate an infitesimal gauge transformation of $G$.
Here index $i = 1,2,3$ corresponds to that of SU(2) generators. 
The explicit forms of $\xi_i^{\alpha}$s for $S^2$ are:
\begin{align}
\xi_1^{\theta} &= \sin \phi ,  \qquad \xi_1^{\phi} = \cot \theta \cos \phi, \nonumber \\
\xi_2^{\theta} &= -\cos \phi ,  \qquad \xi_2^{\phi} = \cot \theta \sin \phi, \nonumber \\
\xi_3^{\theta} &= 0 ,  \qquad \xi_3^{\phi} = -1. 
\end{align} 
The LHSs of Eq~(\ref{symm-con-vec4},\ref{symm-con-vecex}) are infintesimal isometry SU(2)$_I$ transformation and 
the RHSs of those are infintesimal gauge transformation.
%


%
The non-trivial boundary conditions are defined so as to remain the action Eq~(\ref{6Daction}) invariant, 
and are written as
\begin{align}
\label{paripsi}
\psi (x,\pi-\theta,-\phi) &= \gamma_5 P \psi (x,\theta,\phi), \\
\label{pariAmu}
A_{\mu} (x,\pi-\theta,-\phi) &= P A_{\mu}(x,\theta,\phi) P, \\
\label{pariAthe}
A_{\theta} (x,\pi-\theta,-\phi) &= -P A_{\theta}(x,\theta,\phi) P, \\
\label{pariAph}
A_{\phi} (x,\pi-\theta,-\phi) &= -P A_{\phi}(x,\theta,\phi) P, \\ 
\label{boundpsi}
\psi (x,\theta,\phi+2 \pi) &= P' \psi (x,\theta,\phi), \\
\label{boundAmu}
A_{\mu} (x,\theta,\phi+ 2 \pi) &= P' A_{\mu}(x,\theta,\phi) P', \\
\label{boundAthe}
A_{\theta} (x,\theta,\phi+ 2 \pi) &= P' A_{\theta}(x,\theta,\phi) P', \\
\label{boundAph}
A_{\phi} (x,\theta,\phi+2 \pi) &= P' A_{\phi}(x,\theta,\phi) P',
\end{align}
where $P(P')$s act on the representation space of gauge group $G$ and satisfy $P^2=1((P')^2=1)$;
we can take element of $P(P')$ as $\pm 1$. 
%
%
%
%
\subsection{The dimensional reduction and a Lagrangian in four-dimensions}
The dimensional reduction of gauge sector is explicitly carried out by applying the solutions of 
the symmetry condition Eq~(\ref{symm-con-vec4},\ref{symm-con-vecex}).
These solutions are given by Manton~\cite{Manton:1979kb} as 
\begin{align}
\label{kaiAmu}
A_{\mu} &= A_{\mu}(x), \\
\label{kaiAtheta}
A_{\theta} &= -\Phi_1(x), \\
\label{kaiAphi}
A_{\phi} &= \Phi_2(x) \sin \theta - \Phi_3 \cos \theta, \\ 
\label{solW1}
W_1 &= - \Phi_3 \frac{\cos \phi}{\sin \theta}, \\
\label{solW2}
W_2 &= - \Phi_3 \frac{\sin \phi}{\sin \theta}, \\
\label{solW3}
W_3 &= 0,
\end{align}
and satisfy the following constraints:
\begin{align}
\label{kousoku1}
[\Phi_3,A_{\mu}] &= 0, \\ 
\label{kousoku2}
[-i \Phi_3,\Phi_i(x)] &= i \epsilon_{3ij} \Phi_j(x), 
\end{align}
where $\Phi_1(x)$ and $\Phi_2(x)$ are scalar fields, and $-i\Phi_3$ are chosen as generator of U(1)$_I$. 
Note that the $\Phi_3$ term in Eq.~(\ref{kaiAphi}) corresponds to the background gauge field~\cite{background}.
Substituting the solutions Eq~(\ref{kaiAmu})-(\ref{kaiAphi}) into $A_M(X)$ in action Eq~(\ref{6Daction}), 
we can easily integrate coordinates $\theta$ and $\phi$ in 
the gauge sector.
We then obtain a four dimensional action as 
\begin{align}
\label{4d-action}
S_{4D}^{(gauge)} = \int d^4x 
\biggl( 
&- \frac{1}{4g^2} Tr[F_{\mu \nu} F^{\mu \nu}(x)] \nonumber \\
&- \frac{1}{2g^2} Tr[D'_{\mu}\Phi_1(x) D'^{\mu} \Phi_1(x)+D'_{\mu}\Phi_2(x) D'^{\mu} \Phi_2(x)] \nonumber \\
&- \frac{1}{2g^2} Tr[(\Phi_3+[\Phi_1(x),\Phi_2(x)])(\Phi_3+[\Phi_1(x),\Phi_2(x)])]  
\biggr), 
\end{align}
where $D'_{\mu} \Phi = \partial_{\mu} -[A_{\mu},\Phi]$.
The fermion sector of four-dimensional action is obtained by expanding fermions 
in normal modes of $S^2$ and then integrating $S^2$ coordinate in six-dimensional action.
Thus,
 the fermions have massive KK modes which would be a candidate of dark matter.
Generally, the KK modes do not have massless mode because of 
the positive curvature of $S^2$~\cite{A.A.Abrikosov}.
We, however, can show that the fermion components satisfying the following condition have massless mode:
\begin{equation} 
\label{kousoku3}
-i \Phi_3 \psi = \frac{\Sigma_3}{2} \psi.
\end{equation}
Square mass of the KK modes are eigenvalues of square of extra-dimensional Dirac-operator $-i \hat{D}$.
In the $S^2$ case, $-i \hat{D}$ is written as
\begin{align}
\label{dirac}
-i \hat{D} &= -i e^{\alpha a} \Gamma_a D_{\alpha} \nonumber \\ 
&=-i \bigl[ \Sigma_1 (\partial_{\theta} + \frac{\cot \theta}{2} ) 
+ \Sigma_2 (\frac{1}{\sin \theta} \partial_{\phi} + \Phi_3 \cot \theta ) \bigr],
\end{align}
where $\Sigma_i=\mathbf{I}_4 \times \sigma_i$.
Square of $-i \hat{D}$ can be explicitly calculated:
\begin{align}
\label{dirac-square}
(-i \hat{D})^2 = - \bigl[ \frac{1}{\sin \theta} \partial_{\theta} (\sin \theta \partial_{\theta}) 
+ \frac{1}{\sin^2 \theta} \partial_{\phi}^2 +i (2(-i\Phi_3) -  \Sigma_3) \frac{\cos \theta}{\sin^2 \theta} \partial_{\phi} \nonumber \\
-\frac{1}{4} -\frac{1}{4 \sin^2 \theta} + \Sigma_3 (-i\Phi_3 ) \frac{1}{\sin^2 \theta} - (-i\Phi_3)^2 \cot^2 \theta \bigr].
\end{align}
We then act this operator on a fermion $\psi(X)$ which satisfy Eq.~(\ref{kousoku3}), and obtain the reration
\begin{equation}
(-i \hat{D})^2 \psi = -\bigl[\frac{1}{\sin \theta} \partial_{\theta} (\sin \theta \partial_{\theta}) 
+ \frac{1}{\sin^2 \theta} \partial_{\phi}^2 \bigr] \psi.
\end{equation}
The eigenvalues of the RHS operator are less than or equal to zero.
Thus the fermion components satisfying Eq.~(\ref{kousoku3}) have massless mode, while other components 
only have massive KK mode.
Note that the massless mode $\psi_0$ should be independent of $S^2$ coordinates $\theta$ and $\phi$:
\begin{equation}
\label{masslessmode}
\psi_0 = \psi(x).
\end{equation}
The existence of massless fermion may indicate the meaning of the symmetry condition; though the energy density of 
the gauge sector in the appearance of the background fields is higher than 
that of no background fields, since we have massless fermions,
it may consist a ground state as a total in the presence of fermions.
We also note that we could impose symmetry condition on fermions~\cite{Kapetanakis:1992hf,Manton:1981es}. 
In that case, we obtain the massless condition Eq.~(\ref{kousoku3}) from symmetry condition of fermion, 
and 
the solution of symmetry condition is independent 
from $S^2$ coordinate: $\psi=\psi(x)$ 
with no massive KK mode.
Therefore,
we can apply the same discussion for this case as our case  
if we 
only focus on the massless mode in our scheme.



\subsection{A gauge symmetry and particle contents in four-dimensions}

The symmetry conditions and the non-trivial boundary conditions substantially constrain the 
four dimensional gauge group and its representations for the particle contents.
The gauge symmetry and particle contents in four-dimensions 
must satisfy the constraints Eq~(\ref{kousoku1}),(\ref{kousoku2}),(\ref{kousoku3})
and 
be consistent with the boundary conditions Eq~(\ref{paripsi})-(\ref{boundAph}).
We show the prescriptions to identify four-dimensional gauge symmetry and particle contents below.
%


%
First, we show the prescriptions to identify 
gauge symmetry and field components which satisfy the constrants Eq~(\ref{kousoku1}),(\ref{kousoku2}),(\ref{kousoku3}). 
The gauge group $H$ 
that satisfy the constraint Eq~(\ref{kousoku1}) is identified as 
\begin{equation}
\label{H-condition}
H = C_G(U(1)_I)
\end{equation}
where $C_G(U(1)_I)$ denotes the centralizer of U(1)$_I$ in $G$~\cite{Forgacs:1979zs}. 
Note that this implies $G$ $\supset$ $H$ = $H'$ $\times$ U(1)$_I$, where $H'$ is some subgroup of $G$.
%


%
Second, the scalar field components which satisfy the constraints Eq.~(\ref{kousoku2}) are specified by the following 
prescription. 
Suppose that the adjoint representations of SU(2)$_I$ and $G$ are decomposed according to 
the embeddings SU(2)$_I$ $\supset$ U(1)$_I$ and $G$ $\supset$ $H'$ $\times$ U(1)$_I$ as 
\begin{align}
 3( \mathrm{adj} \, \mathrm{SU}(2)) & = (0(\mathrm{adj} \, \mathrm{U}(1)_R)) + (2)+(-2),
  \label{SU(2)-dec}
  \\
  \mathrm{adj} \, G
  & = (\mathrm{adj} \, H)(0)
    + 1(0(\mathrm{adj} \, \mathrm{U}(1))_R)
    + \sum_{g} h_{g}(r_{g}),
  \label{G-dec}
\end{align}
where $h_g$s denote representation of $H'$, and $r_g$s denote U(1)$_I$ charges.
The scalar components satisfying the constraints belong to 
$h_g$s whose corresponding $r_g$s in the decomposition Eq.~(\ref{G-dec}) are $\pm 2$.
%


%
Third, the fermion components which satisfy the constraints Eq.~(\ref{kousoku3}) are determined as follows~\cite{Manton:1981es}.
Let the group U(1)$_I$ be embedded into the Lorentz group SO(2) in such a way that 
the vector representation 2 of SO(2) is decomposed according to SO(2) $\supset$ U(1)$_I$ as 
\begin{equation}
\label{dec-vec}
2= (2)+(-2).
\end{equation}
This embedding specifies a decomposition of the weyl spinor representation $\sigma_6$=4 of SO(1,5) 
according to SO(1,5) $\supset$ SU(2) $\times$ SU(2) $\times$ U(1)$_I$ as
\begin{equation}
\sigma_6 = (2,1)(1) + (1,2)(-1),
\end{equation}
where SU(2) $\times$ SU(2) representations (2,1) and (1,2) correspond to 
left-handed and right-handed spinors, respectively.
We then decompose $F$ according to $G$ $\supset$ $H'$ $\times$ U(1)$_I$ as
\begin{equation}
\label{dec-F}
F = \sum_f h_f(r_f).
\end{equation}
Now the fermion components satisfying the constraints are identified as  
$h_f$s whose corresponding $r_f$s in the decomposition Eq.~(\ref{dec-F}) are (1) for 
left-handed fermions and (-1) for right-handed fermions.
%


%
Finally, we show which gauge symmetry and field components remain in four-dimensions
by surveying the consistency between the boundary conditions Eq.~(\ref{paripsi})-(\ref{boundAph}), 
the solutions Eq.~(\ref{kaiAmu})-(\ref{kaiAphi}),
 and fermion massless mode Eq.~(\ref{masslessmode}).
We then apply Eq~(\ref{kaiAmu})-(\ref{kaiAphi}) and Eq.~(\ref{masslessmode}) to Eq.~(\ref{paripsi})-(\ref{boundAph}), 
and obtain the parity conditions 
\begin{align}
\label{pari-con-Amu}
A_{\mu}(x) &=  P^{(,)}A_{\mu}(x) P^{(,)}, \\
\label{pari-con-sca1}
-\Phi_1(x) &= -P (-\Phi_1(x)) P , \\
\label{pari-con-sca2}
-\Phi_1(x) &= P' (-\Phi_1(x)) P', \\
\label{pari-con-sca3}
 \Phi_2(x)+ \Phi_3 \cos \theta &= -P \Phi_2(x) P+ P \Phi_3 P \cos \theta, \\
 \label{pari-con-sca4}
 \Phi_2(x) - \Phi_3 \cos \theta &= P' \Phi_2(x) P' - P' \Phi_3 P' \cos \theta, \\
 \label{pari-con-psi1}
\psi (x) &=  \gamma^5 P\psi (x), \\
\label{pari-con-psi2}
\psi (x) &= P'\psi (x).
\end{align}
We find that gauge fields, scalar fields and massless fermions in four-dimensions should be even for 
$P A_{\mu} P$ and $P' A_{\mu} P'$; $-P \Phi_{1,2} P $ and $P' \Phi_{1,2} P'$; 
$\gamma_5 P \psi$ and $P' \psi$, respectively. 
$\Phi_3$ always remains since 
it 
is proportional to an U(1)$_I$ generator and commutes with $P(P')$.     
Therefore the particle contents are identified as the components which 
satisfy both the constraints Eq~(\ref{kousoku1}),(\ref{kousoku2}),(\ref{kousoku3}) and 
the parity conditions Eq Eq~(\ref{pari-con-Amu})-(\ref{pari-con-psi2}).
The gauge symmetry remained in four-dimensions can also be identified 
by observing which components of the gauge fields remain.



\section{The SO(12) model}   
\label{SO(12)model}

In this section, we discuss a model based on a gauge group $G$=SO(12) and 
a representation $F$=32 of SO(12) for fermions.
The choice of $G$=SO(12) and $F$=32 is motivated by the study based on CSDR which leads to  
an SO(10) $\times$ U(1) gauge theory with one generation of fermion in four-dimensions~\cite{Chapline:1982wy} 
(for SO(12) GUT see also \cite{Rajpoot:1981it}).



\subsection{A gauge symmetry and particle contents}

First, we show the particle contents in four-dimensions without parities Eq.~(\ref{paripsi})-(\ref{boundAph}).
We assume that U(1)$_I$ is embedded into SO(12) such as 
\begin{equation}
SO(12) \supset SO(10) \times U(1)_I.
\end{equation} 
Thus we identify SO(10) $\times$ U(1)$_I$ as the gauge group which satisfy the constraints Eq~(\ref{kousoku1}), using 
Eq.~(\ref{H-condition}).
We identify the scalar components which satisfy Eq.~(\ref{kousoku2}) 
by decomposing adjoint representation of SO(12):
\begin{equation}
\label{dec66-1}
  SO(12) \supset SO(10) \times U(1)_I: 
  66 = 45(0) +1(0)+ 10(2) + 10(-2).
\end{equation}
According to the prescription below Eq.~(\ref{H-condition}) in sec.~\ref{CSDR}, 
the scalar components 10(2)+10(-2) remains in four-dimensions.
We also identify the fermion components which satisfy Eq.~(\ref{kousoku3}) by decomposing 32 representations of SO(12) as
\begin{equation}
\label{dec32-1}
  SO(12) \supset SO(10) \times U(1)_I: 
  32 = 16(1)+ \overline{16}(-1).
\end{equation}
According to the prescription below Eq.~(\ref{G-dec}) in sec.~\ref{CSDR}, we have the fermion components as 
16(1) for a left-handed fermion and $\overline{16}$(-1) for a right-handed fermion,
 respectively, in four-dimensions.
%


%
Next, we specify the parity assignment of $P(P')$ in order to identify the gauge symmetry and particle contents 
that actually remain in four-dimensions.
We choose a parity assignment so as to break gauge symmetry as 
SO(12) $\supset$ SO(10) $\times$ U(1)$_I$ $\supset$ SU(5)$\times$ U(1)$_X$ $\times$ U(1)$_I$
$\supset$ SU(3) $\times$ SU(2)$_L$ $\times$ U(1)$_Y$ $\times$ U(1)$_X$ $\times$ U(1)$_I$, and to 
maintain Higgs-doublet in four-dimensions.
The parity assignment is written in 32 dimensional spinor basis of SO(12) such as 
\begin{align}
\label{pari32}
SO(12) & \supset  SU(3) \times SU(2)_L \times U(1)_Y \times U(1)_X \times U(1)_I \nonumber \\
32
 = & (3,2)^{(+-)}(1,-1,1)+(\bar{3},2)^{(+-)}(-1,1,-1) \nonumber \\ 
&  + (3,1)^{(--)}(4,1,-1)+(\bar{3},1)^{(--)}(-4,-1,1) \nonumber \\
&  + (3,1)^{(-+)}(-2,-3,-1)+(\bar{3},1)^{(-+)}(2,3,1) \nonumber \\
&  +(1,2)^{(++)}(3,-3,-1)+(1,2)^{(++)}(-3,3,1) \nonumber \\ 
&  + (1,1)^{(--)}(6,-1,1)+(1,1)^{(--)}(-6,1,-1) \nonumber \\
&  +(1,1)^{(-+)}(0,-5,1)+(1,1)^{(-+)}(0,5,-1), 
\end{align}
where e.g. $(+,-)$ means that the parities $(P, P')$ of 
the associated components are (even, odd).
We find the gauge symmetry in four-dimensions by surveying parity assignment for 
the gauge field.
The parity assignments of the gauge field under $A_{\mu}$ $\rightarrow$ $ PA_{\mu}P(P' A_{\mu} P')$ are: 
\begin{align}
\label{pari66-1}
66
 = & (8,1)^{(++)}(0,0,0)+(1,3)^{(++)}(0,0,0)+(1,1)^{(++)}(0,0,0) \nonumber \\
&  +(1,1)^{(++)}(0,0,0)+(1,1)^{(++)}(0,0,0) \nonumber \\
&  + \bigl[(3,2)^{(-+)}(-5,0,0)+ (\bar{3},2)^{(-+)}(5,0,0) \nonumber \\
&  +(3,2)^{(--)}(1,4,0)+ (\bar{3},2)^{(--)}(-1,-4,0) \nonumber \\ 
&  +(3,1)^{(+-)}(4,-4,0)+ (\bar{3},1)^{(+-)}(-4,4,0) \nonumber \\
&  +\underline{(3,1)^{(+-)}(-2,2,2)+ (\bar{3},1)^{(+-)}(2,-2,-2)} \nonumber \\
&  +\underline{(3,1)^{(++)}(-2,2,-2)+ (\bar{3},1)^{(++)}(2,-2,2)} \nonumber \\
&  +\underline{(1,2)^{(--)}(3,2,2)+ (1,2)^{(--)}(-3,-2,-2)} \nonumber \\
&  +\underline{(1,2)^{(-+)}(3,2,-2)+ (1,2)^{(-+)}(-3,-2,2)} \nonumber \\
&  +(1,1)^{(+-)}(6,4,0)+ (1,1)^{(+-)}(-6,-4,0) \bigr].
\end{align}
The components with 
an underline are originated from 
10(2) and 10(-2) of SO(10) $\times$ U(1)$_I$,
which do not satisfy constraints Eq.~(\ref{kousoku1}), 
and hence these components do not remain in four-dimensions. 
Thus we have the gauge field with $(+,+)$ parity components without 
an underline in four-dimensions,
and the gauge symmetry is SU(3) $\times$ SU(2)$_L$ $\times$ U(1)$_Y$ $\times$ U(1)$_X$ $\times$ U(1)$_I$. 
%


%
The scalar particle contents in four-dimensions are determined by 
the parity assignment, under $\Phi_{1,2}$ $\rightarrow$ $-P \Phi_{1,2} P$ and $P' \Phi_{1,2}P'$:
\begin{align}
\label{pari66-2}
66
 = & (8,1)^{(-+)}(0,0,0)+(1,3)^{(-+)}(0,0,0)+(1,1)^{(-+)}(0,0,0) \nonumber \\
&  +(1,1)^{(-+)}(0,0,0)+(1,1)^{(-+)}(0,0,0) \nonumber \\
&  + \bigl[(3,2)^{(++)}(-5,0,0)+ (\bar{3},2)^{(++)}(5,0,0) \nonumber \\
&  +(3,2)^{(+-)}(1,4,0)+ (\bar{3},2)^{(+-)}(-1,-4,0) \nonumber \\ 
&  +(3,1)^{(--)}(4,-4,0)+ (\bar{3},1)^{(--)}(-4,4,0) \nonumber \\
&  +\underline{(3,1)^{(--)}(-2,2,2)+ (\bar{3},1)^{(--)}(2,-2,-2)} \nonumber \\
&  +\underline{(3,1)^{(-+)}(-2,2,-2)+ (\bar{3},1)^{(-+)}(2,-2,2)} \nonumber \\
&  +\underline{(1,2)^{(+-)}(3,2,2)+ (1,2)^{(+-)}(-3,-2,-2)} \nonumber \\
&  +\underline{(1,2)^{(++)}(3,2,-2)+ (1,2)^{(++)}(-3,-2,2)} \nonumber \\
&  +(1,1)^{(--)}(6,4,0)+ (1,1)^{(--)}(-6,-4,0) \bigr]. 
\end{align}
Note that the relative sign for 
the parity assignment of $P$ is different from Eq.~(\ref{pari66-1}), 
and that 
the only underlined parts satisfy the constraints Eq.~(\ref{kousoku2}).
Thus the scalar components in four-dimensions are (1,2)(3,2,-2) and (1,2)(-3,-2,2).
%


%
We find massless fermion contents in four-dimensions, 
by surveying the parity assignment for each components of fermion fields.
We introduce 
two types of left-handed Weyl fermions
that belong to 32 representation of SO(12), which have parity assignment 
$\psi^{(P')}$ $\rightarrow$ $ \gamma_5 P \psi^{(P')}(P' \psi^{(P')})$ 
and $\psi^{(-P')}$ $\rightarrow$ $\gamma_5 P \psi^{(-P')}(-P' \psi^{(-P')})$ respectively.
They have the parity assignment as 
\begin{align}
\label{pari32L-1}
32_L^{(P')}  
 = & \underline{(3,2)^{(--)}(1,-1,1)_L}+(\bar{3},2)^{(--)}(-1,1,-1)_L \nonumber \\
&  +\underline{(\bar{3},1)^{(+-)}(-4,-1,1)_L} + (3,1)^{(+-)}(4,1,-1)_L \nonumber \\
&  +\underline{(\bar{3},1)^{(++)}(2,3,1)_L} + (3,1)^{(++)}(-2,-3,-1)_L \nonumber \\
&  +\underline{(1,2)^{(-+)}(-3,3,1)_L} + (1,2)^{(-+)}(3,-3,-1)_L \nonumber \\ 
&  + \underline{(1,1)^{(+-)}(6,-1,1)_L}+(1,1)^{(+-)}(-6,1,-1)_L \nonumber \\
&  +\underline{(1,1)^{(++)}(0,-5,1)_L}+(1,1)^{(++)}(0,5,-1)_L,  \\
\label{pari32R-1}
32_R^{(P')} 
 = & (3,2)^{(+-)}(1,-1,1)_R+\underline{(\bar{3},2)^{(+-)}(-1,1,-1)_R} \nonumber \\ 
&  +(\bar{3},1)^{(--)}(-4,-1,1)_R + \underline{(3,1)^{(--)}(4,1,-1)_R} \nonumber \\
&  +(\bar{3},1)^{(-+)}(2,3,1)_R + \underline{(3,1)^{(-+)}(-2,-3,-1)_R} \nonumber \\
&  +(1,2)^{(++)}(-3,3,1)_R + \underline{(1,2)^{(++)}(3,-3,-1)_R} \nonumber \\ 
&  + (1,1)^{(--)}(6,-1,1)_R+ \underline{(1,1)^{(--)}(-6,1,-1)_R} \nonumber \\
&  +(1,1)^{(-+)}(0,-5,1)_R+ \underline{(1,1)^{(-+)}(0,5,-1)_R}, 
\end{align}
and 
\begin{align}
\label{pari32L-2}
32_L^{(-P')}  
 = & \underline{(3,2)^{(-+)}(1,-1,1)_L}+(\bar{3},2)^{(-+)}(-1,1,-1)_L \nonumber \\
&  +\underline{(\bar{3},1)^{(++)}(-4,-1,1)_L} + (3,1)^{(++)}(4,1,-1)_L \nonumber \\
&  +\underline{(\bar{3},1)^{(+-)}(2,3,1)_L} + (3,1)^{(+-)}(-2,-3,-1)_L \nonumber \\
&  +\underline{(1,2)^{(--)}(-3,3,1)_L} + (1,2)^{(--)}(3,-3,-1)_L \nonumber \\ 
&  + \underline{(1,1)^{(++)}(6,-1,1)_L}+(1,1)^{(++)}(-6,1,-1)_L \nonumber \\
&  +\underline{(1,1)^{(+-)}(0,-5,1)_L}+(1,1)^{(+-)}(0,5,-1)_L,  \\
\label{pari32R-2}
32_R^{(-P')} 
 = & (3,2)^{(++)}(1,-1,1)_R+\underline{(\bar{3},2)^{(++)}(-1,1,-1)_R} \nonumber \\ 
&  +(\bar{3},1)^{(-+)}(-4,-1,1)_R + \underline{(3,1)^{(-+)}(4,1,-1)_R} \nonumber \\
&  +(\bar{3},1)^{(-+)}(2,3,1)_R + \underline{(3,1)^{(-+)}(-2,-3,-1)_R} \nonumber \\
&  +(1,2)^{(+-)}(-3,3,1)_R + \underline{(1,2)^{(+-)}(3,-3,-1)_R} \nonumber \\ 
&  + (1,1)^{(-+)}(6,-1,1)_R+ \underline{(1,1)^{(-+)}(-6,1,-1)_R} \nonumber \\
&  +(1,1)^{(--)}(0,-5,1)_R+ \underline{(1,1)^{(--)}(0,5,-1)_R}, 
\end{align}
where L(R) means left-handedness(right-handedness) of fermions in four-dimensions, and 
the underlined parts correspond to the components which satisfy constraints Eq.~(\ref{kousoku3}). 
Note the relative sign for parity assignment of $P$ between left-handed fermion and right-handed fermion, 
and that of $P'$ between 32$^{(P')}$ and 32$^{(-P')}$.   
The difference between 32$^{(P')}$ and 32$^{(-P')}$ is allowed because of the bilinear form of the fermion sector.
We thus find that 
the massless fermion components in four-dimensions are one generation of SM-fermions with right-handed neutrino: 
$\{$(3,2)(1,-1,1)$_L$,(3,1)(4,1,-1)$_R$,(3,1)(-2,-3,-1)$_R$,(1,2)(-3,3,1)$_L$,(1,1)(-6,1,-1)$_R$,(1,1)(0,5,-1)$_R$ $\}$.



\subsection{The Higgs sector of the model}

We analyze the Higgs-sector of our model.
The Higgs-sector $L_{\textrm{Higgs}}$ is the last two terms of Eq.~(\ref{4d-action}):
\begin{align}
L_{\textrm{Higgs}} = &- \frac{1}{2g^2} Tr[D'_{\mu}\Phi_1(x) D'^{\mu} \Phi_1(x)+D'_{\mu}\Phi_2(x) D'^{\mu} \Phi_2(x)] \nonumber \\
&- \frac{1}{2g^2} Tr[(\Phi_3+[\Phi_1(x),\Phi_2(x)])(\Phi_3+[\Phi_1(x),\Phi_2(x)])],
\end{align}
where the first term of LHS is the kinetic term of Higgs and the second term gives the Higgs potential.
We then rewrite the Higgs-sector in terms of 
genuine Higgs field in order to analyze it.
%


%
We first note that the $\Phi_i$s are written as 
\begin{equation}
\Phi_i = i \phi_i = i \phi^a_i Q_a,
\end{equation} 
where $Q_a$s are generators of gauge group SO(12), 
since $\Phi_i$s are originated from gauge fields $A_{\alpha}=iA_{\alpha}^a Q_a$;
for the gauge group generator we assume the normalization Tr($Q_aQ_b$)=-2$\delta_{ab}$. 
Note that we assumed the $-i \Phi_3$ as the generator of U(1)$_I$ embedded in SO(12),
\begin{equation}
-i \Phi_3 = Q_I.
\end{equation} 
We change the notation of the scalar fields according to Eq.~(\ref{SU(2)-dec}) such 
that, 
\begin{equation}
\phi_+ = \frac{1}{2} (\phi_1+i \phi_2), \quad
\phi_- = \frac{1}{2} (\phi_1-i \phi_2),
\end{equation}
in order to express solutions of the constraints Eq.~(\ref{kousoku2}) clearly.
The constraints Eq.~(\ref{kousoku2}) is then rewritten as 
\begin{equation}
\label{commutator}
[Q_I,\phi_+ ] = \phi_+, \qquad
[Q_I,\phi_- ] = -\phi_-.
\end{equation}
The kinetic term $L_{KE}$ and potential $V(\phi)$ term are rewritten in terms of $\phi_+$ and $\phi_-$:
\begin{align}
\label{kinetic}
L_{KE} &= -\frac{1}{g^2} Tr[D'_{\mu}\phi_+(x) D'^{\mu} \phi_-(x) ], \\
\label{potential}
V &=  -\frac{1}{2g^2} Tr[Q_I^2-4Q_I[\phi_+,\phi_- ] +4[\phi_+,\phi_- ][\phi_+,\phi_- ] ],
\end{align}
where covariant derivative $D'_{\mu}$ is $D'_{\mu}\phi_{\pm} = \partial_{\mu}\phi_{\pm} - [A_{\mu},\phi_{\pm}]$.
%


%

Next, we change the notation of SO(12) generators $Q_a$ according to decomposition Eq~(\ref{pari66-1}) such 
that 
\begin{align}
\label{generators}
Q_G =  \{ & Q_i , Q_{\alpha}, Q_Y, Q, Q_I, Q_{ax(-500)},Q^{ax(500)} \nonumber \\ 
&  Q_{ax(140)},Q^{ax(-1-40)},Q_{a(4-40)},Q^{a(-440)} \nonumber \\
&  Q_{a(-22-2)},Q^{a(2-22)},Q_{a(-222)},Q^{a(2-2-2)} \nonumber \\
&  Q_{x(322)},Q^{x(-3-2-2)},Q_{x(32-2)},Q^{x(-3-22)} \nonumber \\
&  Q(640),Q(-6-40) \}, 
\end{align} 
where the order of generators corresponds to Eq~(\ref{pari66-1}), index $i=1-8$ corresponds to SU(3) adjoint rep, 
index $\alpha=1-3$ corresponds to SU(2) adjoint rep, index $a=1-3$ corresponds to SU(3)-triplet, and 
index $x=1,2$ corresponds to SU(2)-doublet.
We write $\phi_{\pm}$ in terms of the genuine Higgs field $\phi_x$ which belongs to (1,2)(3,2,-2), such 
that
\begin{align}
\label{scalar}
\phi_+ = \phi_x Q^{x(-3-22)} \\ 
\phi_- = \phi^x Q_{x(32-2)}, 
\end{align}
where $\phi^x=(\phi_x)^{\dagger}$.  
We also write gauge field $A_{\mu}(x)$ in terms of $Q$s in Eq.~(\ref{generators}) as 
\begin{equation}
\label{gauge}
A_{\mu}(x) = i(A_{\mu}^i Q_i+A_{\mu}^{\alpha} Q_{\alpha}+B_{\mu} Q_Y+C_{\mu} Q+E_{\mu} Q_I).
\end{equation}
We then need commutation relations of $Q^{x(-3-22)}$, $Q_{x(32-2)}$, $Q_{\alpha}$, $Q_Y$, $Q$ and $Q_I$
in order to analyze 
the Higgs sector; we summarized them in Table~\ref{commutators}.
\begin{center}
\begin{table}
\begin{tabular}{lll} \hline
& \multicolumn{2}{l} [$Q^{x(-3-22)}$,$Q_{y(32-2)}$] = 
$-\sqrt{\frac{3}{10}}$ $\delta^x_y$ $Q_Y$ + $-\sqrt{\frac{1}{5}}$ $\delta^x_y$ $Q$
+$\delta^x_y$ $Q_I$ +$\frac{1}{\sqrt{2}}$ $(\sigma^*_{\alpha})^x_y$ $Q_{\alpha}$  \\
&
[$Q_{\alpha}$,$Q_x$] = $-\frac{1}{\sqrt{2}}$ $(\sigma_{\alpha})_x^y$ $Q_y$ \qquad \qquad & 
[$Q_{\alpha}$,$Q^x$] = $\frac{1}{\sqrt{2}}$ $(\sigma^*_{\alpha})^x_y$ $Q^y$ \\ 
&
[$Q_x$,$Q_y$]=0 &
[$Q_Y$,$Q^x$]= $-\sqrt{\frac{3}{10}}$ $Q^x$ \\
&
[$Q$,$Q^x$]= $-\sqrt{\frac{1}{5}}$ $Q^x$ &
[$Q_I$,$Q^x$] = $Q^x$ \\ \hline 
\end{tabular}
\caption{commutation relations of $Q^{x(-3-22)}$, $Q_{x(32-2)}$, $Q_{\alpha}$, $Q_Y$, $Q$ and $Q_I$}
\label{commutators}
\end{table}
\end{center}
Finally, we obtain the Higgs sector with genuine Higgs field
by substituting Eq.~(\ref{scalar})-(\ref{gauge}) into Eq.~(\ref{kinetic}, \ref{potential})
and rescaling the fields $\phi \rightarrow g/\sqrt{2} \phi$ and $A_{\mu} \rightarrow g A_{\mu}$, 
and the couplings $\sqrt{2}g=g_2$ and $\sqrt{6/5} g = g_Y$, 
\begin{equation}
L_{Higgs} = |D_{\mu} \phi_x|^2 - V(\phi),
\end{equation}
where the covariant derivative $D_{\mu} \phi_x$ and potential $V(\phi)$ are 
\begin{align}
D_{\mu} \phi_x &= \partial_{\mu} \phi_x + i g_2 \frac{1}{2} (\sigma_{\alpha})_x^y A_{\alpha \mu} \phi_y 
+ i g_Y \frac{1}{2} B_{\mu} \phi_x + i \sqrt{\frac{1}{5}} g C_{\mu} \phi_x - ig E_{\mu} \phi_x , \\
V &= -\frac{2}{R^2} \phi^x \phi_x + \frac{3g^2}{2} (\phi^x \phi_x)^2,
\end{align}
respectively.
Notice that we explicitly write radius $R$ of $S^2$ in the Higgs potential, 
and that we omitted the constant term in the Higgs potential.
We note that the SU(2)$_L$ $\times$ U(1)$_Y$ parts of the Higgs sector has the same form as the SM Higgs sector. 
Therefore we obtain the electroweak symmetry breaking SU(2)$_L$ $\times$ U(1)$Y$ $\rightarrow$ U(1)$_{EM}$. 
The Higgs field $\phi^x$ acquires vaccume expectation value(VEV) as 
\begin{align}
<\phi> &= \frac{1}{\sqrt{2}} \begin{pmatrix} 0 \\ v \end{pmatrix}, \\
v &= \sqrt{\frac{4}{3}} \frac{1}{g R},
\end{align} 
and W boson mass $m_W$ and Higgs mass $m_H$ are given in terms of radius $R$  
\begin{align}
&  m_{W} = g_2 \frac{v}{2} = \sqrt{\frac{2}{3}} \frac{1}{R}, \\
&  m_H = \sqrt{3}g v = \sqrt{4} \frac{1}{R}.
\end{align} 
The ratio between $m_W$ and $m_H$ is predicted 
\begin{equation}
\frac{m_H}{m_W} = \sqrt{6}.
\end{equation}

\section{Summary and discussions}
\label{summary}

We analyzed a gauge theory defined on the six-dimensional spacetime which has 
an $S^2$ extra-space, with the symmetry condition and non-trivial boundary conditions 
and constructed the model based on SO(12) gauge theory.


We first provided the scheme for constructing a four-dimensional theory from a gauge theory
on six-dimensional spacetime which has extra space $S^2$ with the symmetry condition 
of gauge field and the non-trivial boundary conditions.
We showed the prescriptions to identify the gauge field and the scalar field, which
satisfy 
the symmetry condition and the boundary conditions.
A fermion sector of four-dimensional theory
is also obtained by expanding fermions in normal mode and 
integrating 
the $S^2$ coordinates, although explicit form was not shown.
Massive KK modes of fermions then appear in contrast to scalar and gauge field, which would provide a candidate 
of dark-matter.
They may give a rich phenomena in near future collider experiment. 
To discuss these matters, we have to find the eigenvalues of Eq.~(\ref{dirac}).
We leave this in future work.
We also showed that fermions can have massless mode because of 
the existence of a background gauge field.
The fermion components which have massless modes are then determined by the background gauge field and the 
boundary conditions.


Note that by imposing 
the symmetry condition, we can get massless fermions.
It may indicate the meaning of the symmetry condition; though the energy density of 
the gauge sector in the appearance of the background fields is higher than 
that of no background fields, since we have massless fermions,
it may consist a ground state as a total in the presence of fermions.


We then constructed the model based on the SO(12) gauge theory with fermions which lies in a 32 representation of SO(12).
We showed that SU(3) $\times$ SU(2)$_L$ $\times$ U(1)$_Y$ $\times$ U(1)$_X$ $\times$ U(1)$_I$ gauge symmetry is remained 
in four-dimensions, and that the SM Higgs-doublet is obtained without appearance of extra scalar contents.
One generation of SM fermions are successfully obtained by introducing two types of fermions which have 
different parity assignment under $\theta \rightarrow \pi - \theta$.
We also analyzed the Higgs sector that are obtained from gauge sector of the six-dimensional gauge theory.
The electroweak symmetry breaking is then realized and the Higgs mass value is predicted.


To make our model more realistic, there are several challenges such as
eliminating the extra U(1) symmetries 
and
constructing the realistic Yukawa couplings, 
which are the same as other gauge-Higgs unification models.
We, however, can get not only appropriate one-generation fermion fields 
but also Kaluza-Klein modes.
This suggests that we obtain the dark matter candidate in our model.
Thus it is very important to study this model further.
\section*{Acknowledgement}
This work was supported in part 
by the Grant-in-Aid for the Ministry of Education, Culture, Sports, Science, and Technology, 
Government of Japan (No. 19010485, No. 20025001, 20039001, and 20540251).

\appendix

\section{Geometrical quantity on $S^2$}
We summarize the geometrical quantity on $S^2$ such as vielveins $e^a_{\alpha}$, killing vectors $\xi^{\alpha}_a$
and spin connection $R_{\alpha}^{ab}$.
The vielveins are expressed as 
\begin{align}
e^1_{\theta} &= 1, \nonumber \\
e^2_{\phi} &= \sin \theta, \nonumber \\
e^1_{\phi} &= e^2_{\theta} = 0. 
\end{align}
The non-zero components of spin connection are 
\begin{equation}
R^{12}_{\phi} = - R^{21}_{\phi} = -\cos \theta.
\end{equation}

\end{document}